\documentclass[12pt]{article}

\usepackage{graphicx}
\usepackage{scicite}
\usepackage{amsfonts}

\usepackage{times}

\usepackage{caption}

\newenvironment{sciabstract}{%
\begin{quote} \bf}
{\end{quote}}

\newcounter{lastnote}
\newenvironment{scilastnote}{%
\setcounter{lastnote}{\value{enumiv}}%
\addtocounter{lastnote}{+1}%
\begin{list}%
{\arabic{lastnote}.}
{\setlength{\leftmargin}{.22in}}
{\setlength{\labelsep}{.5em}}}
{\end{list}}

\title{Identifying Topological Order by Entanglement Entropy}

\author{Hong-Chen Jiang$^{1,2}$, Zhenghan Wang$^3$, and Leon Balents$^{1,\ast}$ \\
\\
\normalsize{$^1$Kavli Institute for Theoretical Physics, University
of California, Santa Barbara, CA, 93106, U.S.A.}\\
\normalsize{$^2$Center for Quantum Information, IIIS, Tsinghua University, Beijing, 100084, China}\\
\normalsize{$^3$Microsoft Station Q, University
of California, Santa Barbara, CA, 93106, U.S.A.} \\
\normalsize{$^\ast$To whom correspondence should be addressed; E-mail:
 balents@kitp.ucsb.edu.}
}

\date{}

\begin{document}


\baselineskip24pt


\maketitle


\begin{sciabstract}
  Topological phases are unique states of matter incorporating
  long-range quantum entanglement, hosting exotic
  excitations with fractional quantum statistics.  We report a
  practical method to identify topological phases in arbitrary
  realistic models by accurately calculating the Topological
  Entanglement Entropy (TEE) using the Density Matrix Renormalization
  Group (DMRG).  We argue that the DMRG algorithm
  naturally produces a minimally entangled state, from amongst the
  quasi-degenerate ground states in a topological phase.
  This proposal both explains the success of this method, and the
  absence of ground state degeneracy found in prior DMRG sightings of
  topological phases.  We demonstrate the effectiveness of the
  calculational procedure by obtaining the TEE for several microscopic
  models, with an accuracy of order $10^{-3}$ when the circumference of the
cylinder is around ten times the correlation length.  As an example,
we definitively show the ground state of the quantum
  $S=1/2$ antiferromagnet on the kagom\'e lattice is a topological spin liquid, and
  strongly constrain the full identification of this phase of matter.
\end{sciabstract}


Theory has shown that quantum ground states may exhibit distinct
patterns of {\sl long-range entanglement}, which provides the most
basic categorization of quantum phases of matter, more fundamental
than Landau's symmetry breaking paradigm.  The simplest and most
robust long range entangled states, which have a full spectral gap,
comprise ``topological phases'', which host topological order.  Much
recent interest in topological phases is due to the prospect of
utilizing them to construct an inherently fault tolerant quantum
computer\cite{Kitaev2003,Nayak2008}.  Topological phases are also of
basic scientific interest for their many unique properties,
especially their ability to support exotic excitations with
fractional and even non-abelian quantum statistics.

Crystalline ``Mott'' insulators with unpaired electron spins have long
been considered likely candidates for long range entangled states,
epitomized in this context by Anderson's resonating valence bond\cite{AndersonRVB}
wavefunction for a ``quantum spin liquid''\cite{Balents2010}.  Because this particular
state is non-magnetic, the lack of magnetic order has been widely taken as
a definition of a quantum spin liquid.  However, defining what a
quantum spin liquid {\sl isn't} has little utility, and is especially
unhelpful in the theoretical search for these phases.  Instead, a {\sl
  positive} definition of a quantum spin liquid which can be tractably
tested in realistic models is sorely needed.

In principle, such a positive definition is provided for topological
phases (and hence those quantum spin liquids with topological order)
by the Topological Entanglement Entropy (TEE) introduced by
Kitaev-Preskill\cite{Kitaev2006} and Levin-Wen\cite{Levin2006}.
Unfortunately, the formulations in Refs.\cite{Kitaev2006,Levin2006}
suffer from severe finite-size corrections due to lattice scale
effects, greatly hindering their application.  We report here a
practical and extremely simple scheme to numerically calculate the
TEE, and thereby identify topological order.  Our method consists
simply of using the Density Matrix Renormalization
Group\cite{White1992,Stoudenmire2011} (DMRG) to calculate the usual
entanglement entropy for the division of a cylinder into two equal
halves by a flat cut, and extracting the TEE from its asymptotic,
large circumference limit (see below).  We argue that this method
actually works, despite potential complications known
theoretically,\cite{Dong2008,Zhang2011} due to a subtle ground state
selection mechanism built into the DMRG algorithm.  The approach is
tested here on a variety of lattice models, and then applied
successfully to the physically realistic quantum spin
$S=\frac{1}{2}$ anti-ferromagnetic Heisenberg $J_1$-$J_2$ model on
the kagom\'e lattice. By extracting an accurate TEE, we identify a
quantum spin liquid state with topological order for the first time
in a physically realistic $SU(2)$-invariant lattice model.  We
emphasize that the TEE provides {\sl positive}, ``smoking gun''
evidence for a topological quantum spin liquid, and excludes any
topologically trivial states, regardless of possible complex or
subtle broken symmetries.  The {\sl value} of the TEE also greatly
restricts the possible topological quantum field theories which
fully describe the topological order.  We return to this at the end
of this paper.

The TEE is derived from the bipartite von Neumann entanglement
entropy, which is defined by dividing a system into two subsystems,
$A$ and $B$, which together comprise the full system.  The
entanglement entropy associated to this partition is defined from
the reduced density matrix, $\rho_A = {\rm Tr} \left(
|0\rangle\langle 0|\right)$, where $|0\rangle$ is a ground state,
according to $S(A)=-{\rm Tr} [\rho_A {\ln}(\rho_A)]$.  It has the
duality property $S(A)=S(B)$.  According to the seminal works of
Kitaev-Preskill\cite{Kitaev2006} and Levin-Wen\cite{Levin2006}, the
entanglement entropy of a partition of a two dimensional system
where A is a disk-like region with a {\sl smooth} boundary (the
``entanglement surface'') of length $\ell$ scales as
\begin{eqnarray}
  S(A)=\alpha \ell -\gamma + \cdots,\label{Eq:EntropyDisk}
\end{eqnarray}
where the ellipsis represents terms that vanish in the limit $\ell
\rightarrow \infty$.   The coefficient $\alpha$, due to short
distance physics near the boundary, is non-universal.  The term
$\gamma$ is the topological entanglement entropy (TEE)---a universal
additive constant characterizing the long-range entanglement in the
ground state which can be quantified as $\gamma=\ln D$, where $D$ is
the total quantum dimension of the
medium\cite{Kitaev2006,Levin2006}.

Note that $\gamma$ is {\sl sub-dominant} to the $\alpha \ell$ term,
arising from {\sl short-range entanglement}.  As a consequence, it is
non-trivial to extract.  Moreover, for real lattice systems, it is not
obvious how to define $\ell$ on the lattice, nor is it obvious what
qualifies as a ``smooth" boundary.    These two ambiguities are
particularly challenging.


In Refs.\cite{Kitaev2006,Levin2006}, complex prescriptions were
proposed to remove the short-range contributions and extract the TEE
from measurements on planar partitions.  In our much simpler scheme,
we study a single partition, defined by a straight cut normal to a
cylinder which divides it in half, and extract the TEE using
Eq.~(\ref{Eq:EntropyDisk}) with $\ell=L_y$, the circumference of the
cylinder.  This approach minimizes errors due to subtractions of many
large numbers, and also minimizes finite size corrections due to
short-range entanglement, as we now argue.

For the cylindrical case, we expect such finite size corrections to be
of order $e^{-L_y/\xi}$.  In the Kitaev-Preskill and Levin-Wen
formulations, the corrections are much larger.  There, to obtain the
TEE, the entropy is calculated for several disk-like planar
partitions, and corner contributions are cancelled by forming a linear
combination of the results.  However, the complicated shape of the
planar partitions involved means that the smallest spatial features of
the partition are several times smaller than the overall system width.
For instance, in the Levin-Wen formulation, the smallest features
(size $d$) are {\sl at least} four times smaller than the linear width
of the system assuming periodic boundary conditions, so that $L \geq
4d$, a conservative estimate.  Corrections to
Eq.~(\ref{Eq:EntropyDisk}) should be expected to be of order
$e^{-d/\xi} \geq e^{-L/(4\xi)}$.  Thus to obtain similar performance
to that of the cylindrical cut, even assuming no additional errors are
introduced by the subtractions of different entropies, requires a {\sl
  linear} system size at least four times larger in the Levin-Wen
case.  This means at least 16 times as many spins, and given the
exponential growth of the Hilbert space with the number of quantum
degrees of freedom, this is a very costly increase.  Indeed, attempts
to implement the Kitaev-Preskill and Levin-Wen protocols in
simulations have shown them to be very challenging
numerically\cite{Furukawa2007,Isakov2011}.

A potential complication of our method is that the ground state on a
cylinder is expected to have a degeneracy in a topological phase in
the thermodynamic limit, and the TEE for the cylindrical cut can
depend upon {\sl which} ground state the TEE is measured
in\cite{Dong2008,Zhang2011}.   In Ref.~\cite{Zhang2011}, it has been
shown, however, that the TEE for the $\mathbb{Z}_2$ spin liquid is
bounded above by the universal value $\gamma=\ln2$, and below by
zero.  Moreover, in general the universal value is achieved for
so-called Minimal Entropy States (MES's)\cite{Zhang2011}, which
correspond to states in which a quasiparticle is definitely
contained within the region A (or B).  For the $\mathbb{Z}_2$ spin
liquid, the MES's are the states with a {\sl
  definite} $\mathbb{Z}_2$ magnetic flux through the cylinder,
i.e. the vison or no-vison eigenstates.

We suggest, based on numerical evidence, that {\sl the DMRG
  systematically finds a MES}.  This is perhaps natural since the DMRG
prefers low entanglement states\cite{Stoudenmire2011}. Note though that
the {\sl absolute} ground state of a finite system is dependent upon
microscopic details, and is expected to vary with the aspect
ratio ($L_x/L_y$) of the cylinder\cite{Jiang2011}.  For a physical
Hamiltonian without fine tuning, the absolute ground state becomes a MES
in the ``long'' cylinder limit, where $L_x/L_y$ is larger than some
critical value (which depends on microscopic details, but is order one
generically)\cite{Jiang2011}.  However, we contend that the DMRG
preferentially finds the MES even when it is not the absolute ground
state.  Evidence for this is given below in the toric code model, where
the MES can be explicitly identified.  The fact that we obtain the
universal value of the TEE, independent of the system's aspect ratio, for
several other models, also supports this conclusion.

\begin{figure}
\centerline{\includegraphics[height=4.4in,width=3.4in]{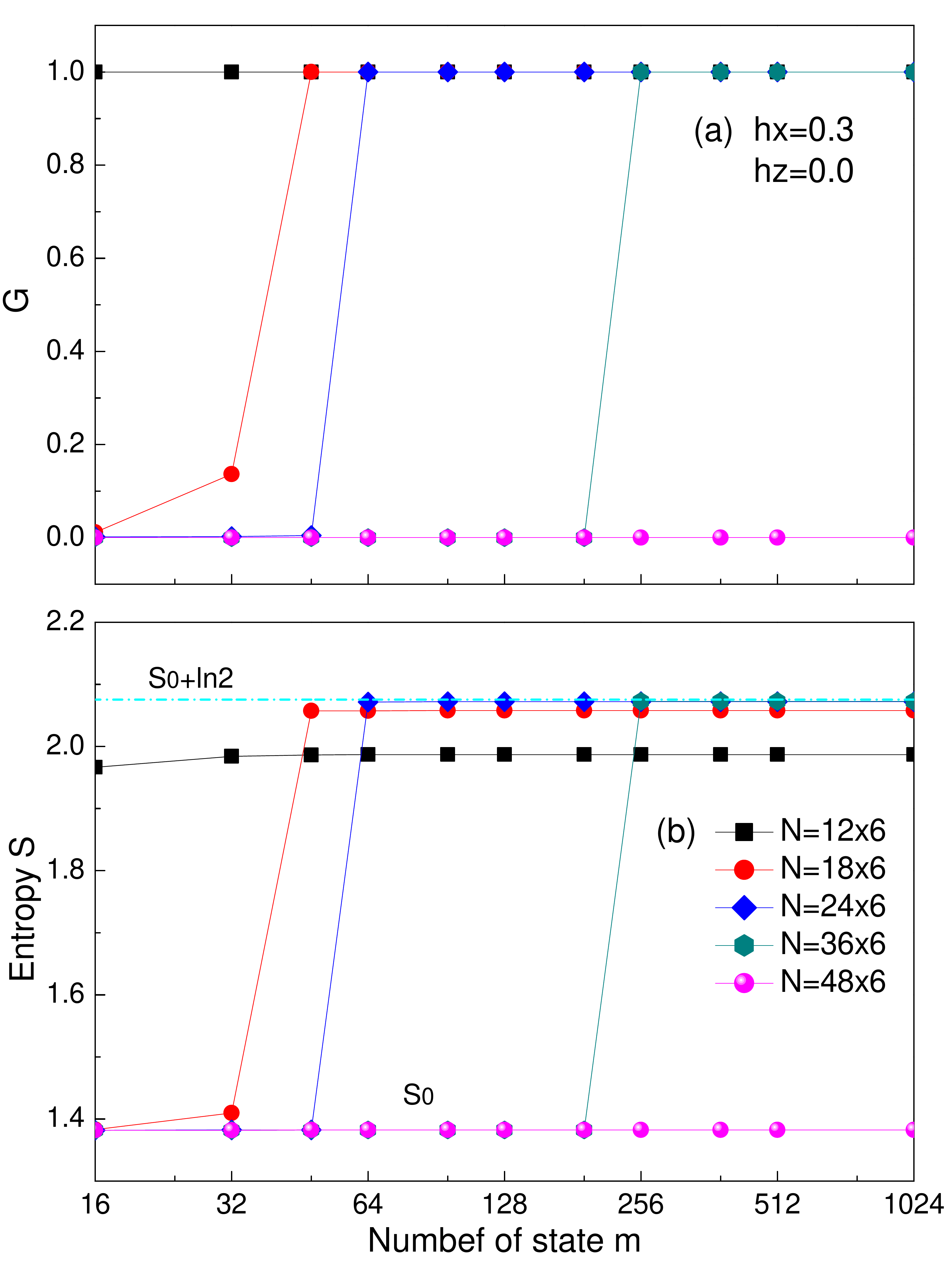}}%
\caption{Evidence that the DMRG favors MES's. In (a) the
  electric field parity $\langle G\rangle$ and (b) the entanglement
  entropy are plotted versus number of states $m$ for the toric code
  with $h_x=0.3$ and $h_z=0$, for several system sizes.  We see that for
  fixed small system size, at smaller $m$ the average parity $\langle
  G\rangle$ is approximately zero and the entanglement is reduced, while
  at large $m$ a definite parity eigenstate is found with $G=1$, and the
  entanglement is increased.  The jumps in the two quantities coincide,
  signaling a transition from a MES to an absolute Hamiltonian
  eigenstate.  The number of states $m$ needed to converge to the
  absolute ground state increases rapidly with $L_x$.  For larger
  systems than shown, a MES with $\langle G\rangle \approx 0$ is
  found for all accessible values of $m$. }
\label{Fig:MES}
\end{figure}
\begin{figure}
\centerline{\includegraphics[height=4.4in,width=3.4in]{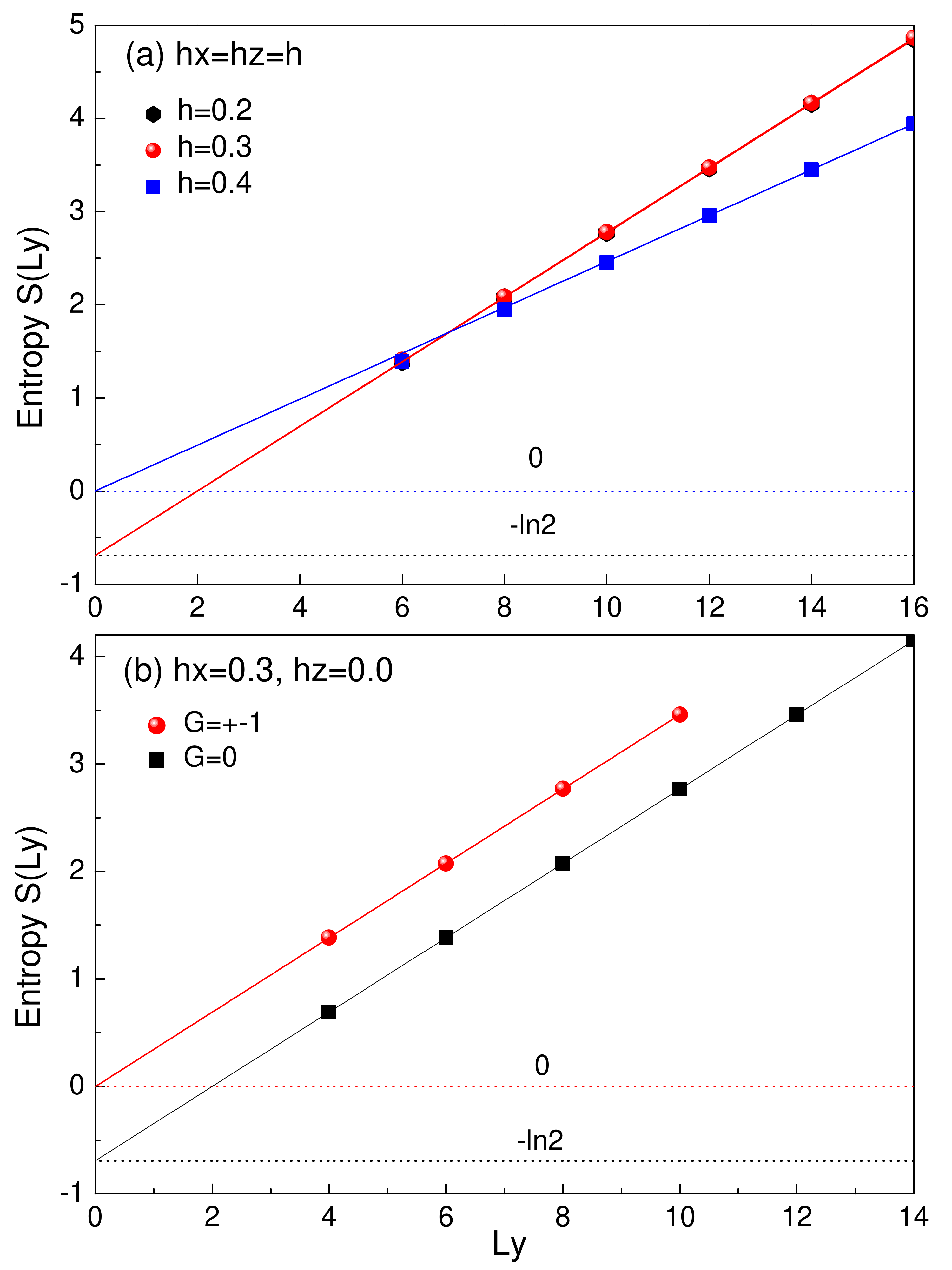}}%
\caption{The von Neumann entropy $S(L_y)$ for the toric code model in
  magnetic fields. In (a), we show $S(L_y)$ with $L_y=4\sim 16$ at
  $L_x=\infty$ for symmetric magnetic fields at $h_x=h_z=h=0.2$, $0.3$
  and $0.4$. By fitting $S(L_y)=aL_y-\gamma$, we get
  $\gamma=0.693(1)$, $0.691(4)$ and $0.001(5)$, respectively. In (b),
  we consider the pure electric case, $h_x=0.3,h_z=0$, and compare
  $S(L_y)$ in the MES obtained in the large $L_x$ limit (black
  squares) to that of the absolute ground state from systems of
  dimensions $L_x\times L_y=20\times 4, 24\times 6, 24\times
  8,24\times 10$ (red circles). Extrapolation shows that the MES has the universal TEE,
  while the absolute ground state has zero
  TEE.} \label{Fig:ToricCodeModel}
\end{figure}

We turn now to the toric code model, which is well known -- see the
Supplementary material for details of the definition.  It can be
considered as a model of fluctuating discrete ``electric'' and
``magnetic'' fields.  To observe the ground state selection, we first
consider an applied field $h=h_x\neq 0$ which is purely electric,
$h_z=0$. Then the operator $G$ (defined in the Supp. Mat.), which
measures the parity of the number of electric field loops winding
around the cylinder, commutes with the Hamiltonian, so the energy
eigenstates must also be eigenstates of $G=\pm 1$.  Topological order
implies that there are two such states with $G=+1$ and $G=-1$, with
exponentially close energies. The MES's, however, are not $G$ (or
energy) eigenstates, but rather the superpositions $|\pm\rangle =
(|G=1\rangle \pm |G=-1\rangle)/\sqrt{2}$, for which $\langle
\pm|G|\pm\rangle=0$.  The $|\pm\rangle$ states correspond to states
with or without magnetic flux through the cylinder. Measurements of
$\langle G\rangle$ and $S$, Figure~\ref{Fig:MES}, show that the DMRG
preferentially selects a MES for larger systems, and that the number
of states $m$ necessary to converge to the absolute ground state (with
larger entanglement and zero TEE -- see
Figure~\ref{Fig:ToricCodeModel}b) grows very rapidly with system size.

The origin of the topological contributions to the entanglement
entropy sheds light on this behavior in the $\mathbb{Z}_2$ case of
interest.  First, there is a {\sl reduction} of entropy due to the
constraint that electric field loops always cross the entanglement
surface an even number of times.  This reduction is precisely the TEE,
and this physics is included once entanglement on the scale of $L_y$
is taken into account.  Second, in the case where the absolute ground
state is not a MES, there is an {\sl increase} of entropy due to the
{\sl global} constraint on the number of electric field lines winding
the {\sl entire} cylinder.  To take this into account, the DMRG must
fully converge the entanglement of the opposite ends of the system,
which are extremely far separated on the ``snaking'' DMRG path.  This
global entanglement does not converge for larger systems, in which
case the DMRG produces states described by a Schmidt decomposition in
which the left and right halves of the system have {\sl uncorrelated}
electric field winding parities.  Such a state is a MES.

We next consider the toric code model in symmetrically applied fields,
$h_x=h_z=h$, for which the absolute ground state is not obvious.
Figure~\ref{Fig:ToricCodeModel} shows the entanglement entropy in this
case.  This model was previously
shown\cite{Trebst2007} to have a quantum phase
transition between the $\mathbb{Z}_2$ phase for $h<h_c\approx 0.34$
and a trivial phase for $h>h_c$.  The extrapolated TEE following our
protocol indeed very well approximates the universal value $\gamma =
\ln2 = 0.69314\ldots$ for $h<h_c$; even for $h=0.3$, relatively close
to the quantum phase transition, we obtain $\gamma=0.691(4)$, which is
accurate to a fraction of a percent.  For $h>h_c$, we obtain
$\gamma=0$, as expected, with a numerical uncertainty of order
$10^{-3}$.  Similar results are obtained for a variety of aspect
ratios and values of the perturbing fields.

We apply the method to the spin-$1/2$ Heisenberg model on the kagom\'e
lattice, for which compelling but indirect evidence for a gapped
quantum spin liquid has been recently obtained by extensive DMRG
studies\cite{Jiang2008,White2011Kagome,White2012}. We consider the
model with both first and second-neighbor interactions,   whose
Hamiltonian is
\begin{eqnarray}
H &=&J_1\sum_{\langle ij\rangle}\textbf{S}_i\cdot
\textbf{S}_j+J_2\sum_{\langle\langle
ij\rangle\rangle}\textbf{S}_i\cdot
\textbf{S}_j,\label{Eq:KagomeJ1J2Model}
\end{eqnarray}
where $\textbf{S}_i$ is the spin operator on site $i$, and $\langle
ij\rangle$ ($\langle\langle ij\rangle\rangle$) denotes the nearest
neighbors (next nearest neighbors). In the numerical simulation, we
set $J_1=1$ as the unit of energy.  The most recent DMRG
studies\cite{White2012} show that the $J_2=0$ point is near the edge
of a substantial spin liquid phase centered near $J_2=0.05-0.15$.

\begin{figure}
\centerline{\includegraphics[height=2.4in,width=3.4in] {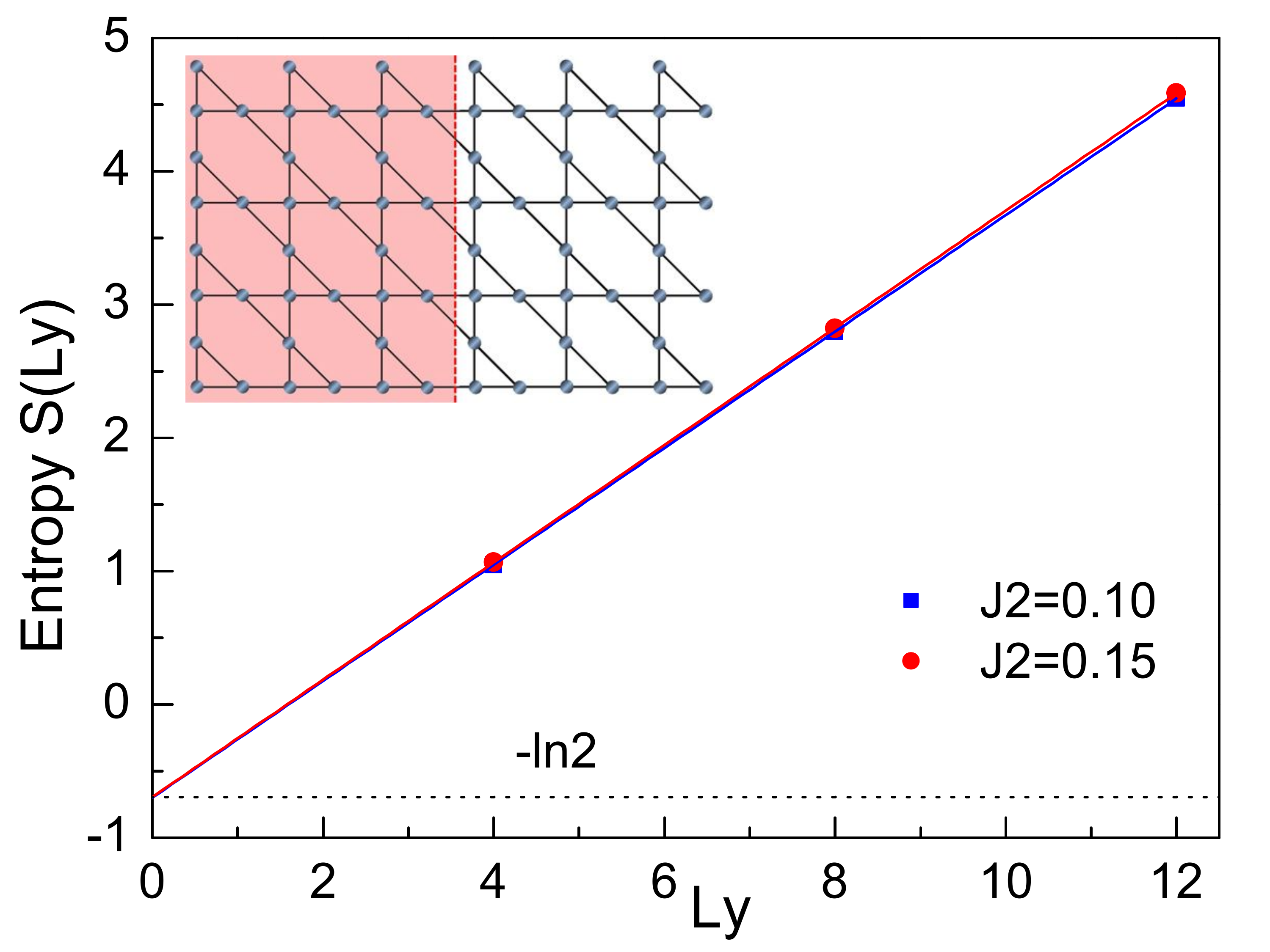}}%
\caption{The entanglement entropy $S(L_y)$ of the kagom\'e J$_1$-J$_2$
model in Eq.(\ref{Eq:KagomeJ1J2Model}), with $L_y=4\sim 12$ at
$L_x=\infty$. By fitting $S(L_y)=aL_y-\gamma$, we get
$\gamma=0.698(8)$ at $J_2=0.10$, and $\gamma=0.694(6)$ at
$J_2=0.15$. Inset: kagom\'e lattice with $L_x=12$ and $L_y=8$.}
\label{Fig:KagomeJ1J2Model}
\end{figure}

We take the kagom\'e lattice with periodic boundary conditions along
a bond direction, drawn vertically in the inset of
Figure~\ref{Fig:KagomeJ1J2Model}, and the unit of length equal to
the nearest-neighbor distance.  The results for the entanglement
entropy for $J_2=0.10$ and $0.15$ are shown in
Figure~\ref{Fig:KagomeJ1J2Model} with correlation length around
one-lattice spacing for both spin-spin and dimer-dimer correlation
functions.  We see that a linear fit using data for $L_y=4\sim 12$
using Eq.(\ref{Eq:EntropyDisk}) gives $\gamma=0.698(8)$ at
$J_2=0.10$ and $\gamma=0.694(6)$ at $J_2=0.15$, both within one
percent of $\ln 2=0.693$.  This proves definitively that this phase
is a topological spin liquid, and determines the quantum dimension
$D=2$, very consistent with a $\mathbb{Z}_2$ state.



We have shown that the TEE can be calculated to an accuracy of order
$10^{-3}$ when $L_y$ is $\sim$10 times the correlation length (see the
Supporting Online Material for some additional tests).  Our result
provides a ``smoking gun'' test for a topological spin liquid.  It also
explains the puzzling absence of topological degeneracy in recent DMRG
results which otherwise support a $\mathbb{Z}_2$ spin liquid
state\cite{Jiang2011,White2011Kagome}, since we have shown that the DMRG
is systematically biased to find just {\sl one} of the ground
states. The TEE does not fully determine the nature of the topological
phase. Fortunately, for a given $D$, there are only finitely many
distinct topological phases, and for small values of $D$, a complete
classification of all topological phases is known\cite{RSW2009}.  Other
constraints such as time-reversal symmetry (if present) further constrain
the possible topological order.  For example, there are only two
time-reversal invariant phases consistent with $D=2$, found here for the
kagom\'e Heisenberg model: the $\mathbb{Z}_2$ phase, and a doubled
semion phase.  It will be interesting to develop methods to distinguish
these in the future, and to calculate the topological ground state
splitting.  Identifying topological order by combining theoretical
classification results with numerical simulation is a major step in the
development of a post-Landau paradigm for classifying quantum phases of
matter.

\bibliographystyle{Science}



\begin{scilastnote}
\item We thank Tarun Grover and Ashvin Vishwanath for a helpful
  explanation of their work, and Steve White for helpful
  discussions. H.C.J. thanks Hong Yao for collaboration on related projects.
  This work was supported by the NSF through
  grant DMR=0804564 (L.B.), the NSF MRSEC Program under DMR 1121053,
the NBRPC (973 Program) 2011CBA00300 (2011CBA00302), and
  benefitted from the facilities of the KITP, supported by NSF
  PHY05-51164.

\item {\bf Supporting Online Material}\\
www.sciencemag.org\\
Materials and Methods\\
Figures S1, S2, S3\\
References (19-25)
\end{scilastnote}


\clearpage


\clearpage

\appendix

\noindent {\huge {\bf Supporting Online Material}}


\section*{Materials and Methods}
\label{sec:materials-methods}

\renewcommand{\thefigure}{S\arabic{figure}}
\setcounter{figure}{0}
\renewcommand{\theequation}{S\arabic{equation}}
\setcounter{equation}{0}

Here we test our method on a variety of lattice models whose topological
order is known.

\section{Toric-code model in magnetic fields}%
The toric code model\cite{Kitaev2003} with an applied magnetic field is given by
\begin{eqnarray}
H=-J_s\sum_s A_s - J_p\sum_p B_p - h_x\sum_i\sigma^x_i -
h_z\sum_i\sigma^z_i,\label{EqS:ToricCodeModel}
\end{eqnarray}
where $\sigma^x_i$ and $\sigma^z_i$ are Pauli matrices, and
$A_s=\Pi_{i\in s}\sigma^x_i$, $B_p=\Pi_{i\in p}\sigma^z_i$.
Subscripts $s$ and $p$ refer to, respectively, vertices and plaquettes of
a square lattice, whereas $i$ runs over all bonds where spin degrees
of freedom are located. Without magnetic field, i.e., $h_x=h_z=0$,
the pure toric code model can be solved exactly\cite{Kitaev2003},
and the ground state has $\mathbb{Z}_2$ topological order with total quantum
dimension $D=2$. On the torus the ground state is $4$-fold degenerate.
All elementary excitations are gapped and characterized by
eigenvalues $A_s=-1$ (a $\mathbb{Z}_2$ charge on site $s$) and $B_p=-1$ (a
$\mathbb{Z}_2$ vortex on plaquette $p$). 
When turning on the magnetic field, the model cannot be solved
exactly anymore. Previous
studies\cite{Trebst2007,Vidal2008,Tupitsyn2008} show that the
$\mathbb{Z}_2$ topological phase remains stable and robust until the
magnetic fields are large enough, where the system crosses the
transition from the topological phase to the trivial one.
Specifically, such a phase transition takes place at the critical
magnetic field $h_c=0.34$ along the symmetric line $h_x=h_z=h$.

\begin{figure}
\centerline{\includegraphics[height=2.4in,width=3.4in]{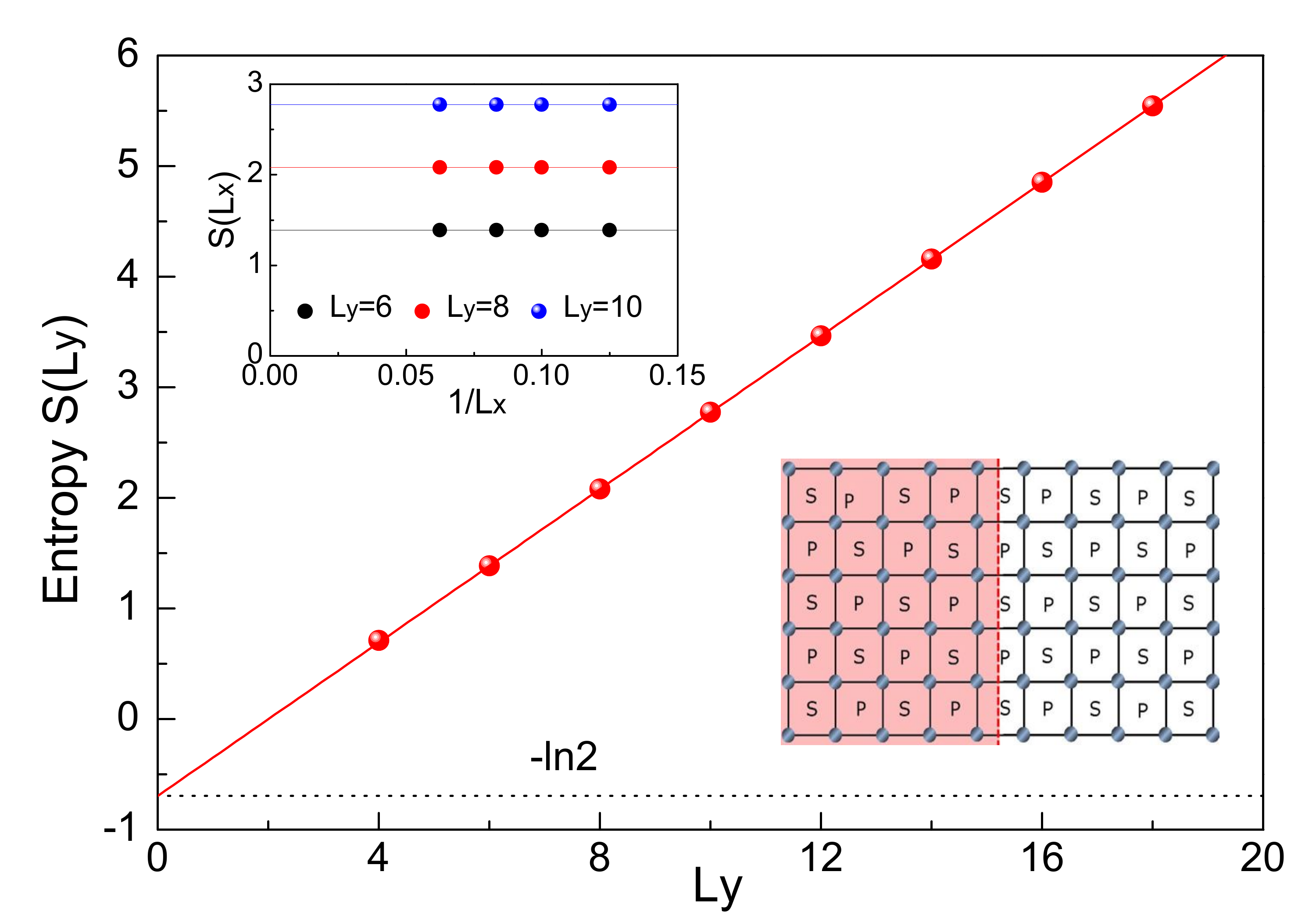}}%
\caption{The von Neumann entropy for the pure toric code model and
that in magnetic fields in Eq.(\ref{EqS:ToricCodeModel}). The von
Neumann entropy $S(L_y)$ for the pure Toric-Code model with
$L_y=4\sim 20$ at $L_x=\infty$. By fitting $S(L_y)=aL_y-\gamma$, we
get $\gamma=0.693147(1)$. Inset: (Upper) the entanglement entropy
$S(L_x)$ as a function of $L_x$ for different $L_y$; and (Lower)
Square lattice with $L_x=10$ and $L_y=6$. Here $S$ represents the
star operator $A_s$, while $P$ represents the plaquette operator
$B_p$.} \label{FigS:ToricCodeModel}
\end{figure}

For the DMRG simulation, we consider an equivalent square lattice,
where the spin operators $\sigma^x$ and $\sigma^z$ sit on the sites
instead of the bonds. Therefore, the star operator $A_s$ and the
plaquette operator $B_p$ of the original lattice now sit on
alternating plaquettes in the equivalent square lattice, as shown in
Figure~\ref{EqS:ToricCodeModel}, labeled as $S$ and $P$,
respectively. Note that on this equivalent square lattice, there are
an even number of dangling spins within each plaquette at the open
edges. For the pure toric-code model with cylinder boundary
condition, the first $2^{L_y/2-1}$ eigenvalues of the reduced
density matrix $\rho_A$ are degenerate and equal to $1/2^{L_y/2-1}$,
while all the other eigenvalues are zero. This allows us to study a
quite large system with width up to $L_y=20$ easily. As shown in
Figure~\ref{FigS:ToricCodeModel}, after fitting the entanglement
entropy using Eq.(\ref{Eq:EntropyDisk}) for $L_x=\infty$, we get a
nonzero topological entanglement entropy $\gamma=0.693147(1)$, which
is equal to the expected value $\gamma=\ln2$ with amazingly small
numerical error $10^{-6}$.

After turning on the magnetic field, the degeneracy of entanglement
spectrum is lifted, and the correlation length $\xi$ becomes finite.
Our results show that even very close to the phase transition point
(e.g., $h_c=0.34$ along the symmetric line), we can still get a very
accurate TEE $\gamma$.  For example, the correlation length $\xi\sim 1$
lattice spacing at $h=0.30$, the resulted topological entanglement
entropy $\gamma=0.691(4)$ is still quite accurate with an error around
$10^{-3}$. These results show that a nonzero TEE is obtained throughout
the topologically ordered phase. On the contrary, the topological
entanglement entropy $\gamma$ is zero in the trivial phase where
$h>h_c$. For example, $\gamma=0.001(5)$ at $h=0.40$. Therefore, our
method allows us to unambiguously extract the non-zero topological
entanglement entropy $\gamma$ if and only if the toric-code model is in
a topologically ordered phase.

In the case of a purely electric perturbation, $h_z=0$, $h_x=h \neq 0$,
two loop operators commute with $H$.  Specifically, these are
\begin{eqnarray}
  \label{eq:1}
  G & = & G_y = \prod_{x=1}^{L_x} \sigma^x_{x,y}, \\
  G_x & = & \prod_{y=1}^{L_y} \sigma^x_{x,y}.
\end{eqnarray}
In the low energy sector where $A_s=+1$ for all $s$, $G_y$ is
independent of $y$ and $G_x$ is independent of $x$.  By construction,
$G$ and $G_x$ have eigenvalues $\pm 1$.  The operator $G_x$ probes the
presence or absence of an electric particle at the end of the cylinder.
This degree of freedom is not associated to the ground state degeneracy,
and indeed we find $G_x=+1$ always in our numerics.  The operator $G$
counts the parity of the number of electric flux lines winding around
the cylinder, and {\sl does} operate in the topologically degenerate
subspace.  Physically, eigenstates of $G$ are equal weight
superpositions of the vison and no-vison eigenstates, which are the
MES, as discussed in the main text.  For $h_z=0$, the energy eigenstates
must also be eigenstates of $G$, and the splitting between them is
expected to be expontially small in $L_x$.

\section{Fractional quantum Hall model}%

We next consider the so-called Haldane model\cite{Haldane1988} on the
honeycomb lattice filled with hard-core bosons:%
\begin{eqnarray}
H&=&-t^\prime \sum_{\langle\langle
rr^\prime\rangle\rangle}\left[b^\dagger_{r^\prime}b_{r}e^{i\phi_{r^\prime
r}}+\rm H.c.\right]\\%
&-&t\sum_{\langle r
r^\prime\rangle}\left[b^\dagger_{r^\prime}b_{r}+\rm H.c.\right]
-t^{\prime\prime}\sum_{\langle\langle\langle r
r^\prime\rangle\rangle\rangle}\left[b^\dagger_{r^\prime}b_{r}\rm
+H.c.\right],\nonumber \label{EqS:HoneycombFQHE}
\end{eqnarray}
where $b^\dagger_{r}$ creates a hard-core boson at site $r$,
$\langle\cdots\rangle$, $\langle\langle\cdots\rangle\rangle$, and
$\langle\langle\langle\cdots\rangle\rangle\rangle$ denote the
nearest-neighbor, the next-nearest-neighbor, and the
next-next-nearest-neighbor pairs of sites, respectively. In
Ref.\cite{Wang2011}, the authors have systematically studied this model
using exact diagonalization, providing convincing evidence showing that
the ground state (with parameters $t^\prime=0.6t$,
$t^{\prime\prime}=-0.58t$, and $\phi=0.4\pi$) is a $1/2$ bosonic FQH
state with two-fold ground state degeneracy on the torus. Such a $1/2$
FQH state has nontrivial semion topological
order\cite{Kitaev2006,Levin2006}, with total quantum dimension
$D=\sqrt{2}$.

\begin{figure}
\centerline{\includegraphics[height=2.4in,width=3.4in]{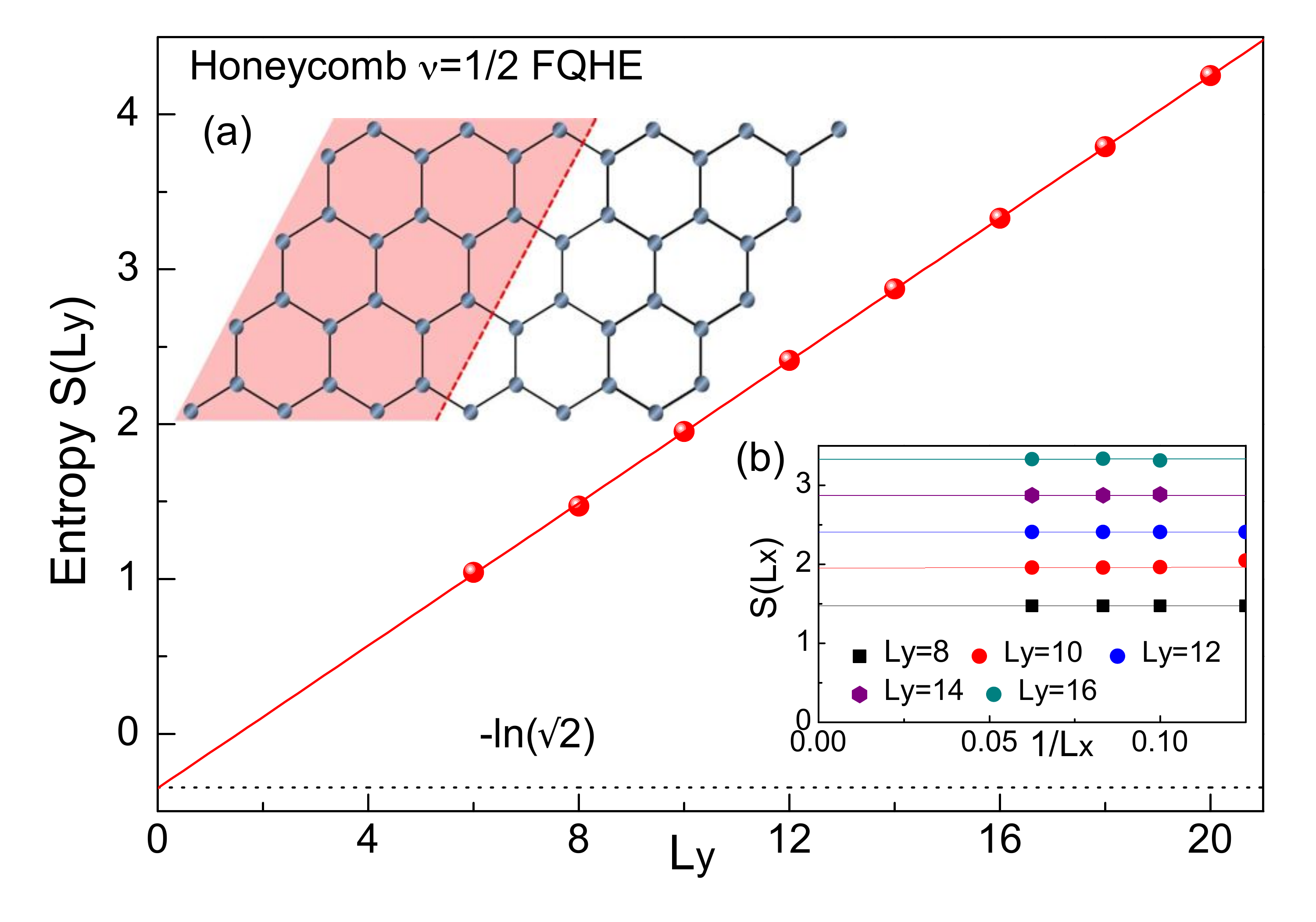}}%
\caption{The entanglement entropy $S(L_y)$ of the Honeycomb Haldane
model in Eq.(\ref{EqS:HoneycombFQHE}), with $L_y=6\sim 20$ at
$L_x=\infty$. By fitting $S(L_y)=aL_y-\gamma$, we get
$\gamma=0.349(5)$. Inset: (a) Honeycomb lattice with $L_1=6$ and
$L_2=4$. Here the system width $L_y=2L_2$, and system length
$L_x=2L_1$. (b) The entanglement entropy $S(L_x)$ as a function of
$L_x$ for different $L_y$.} \label{FigS:HoneycombFQHE}
\end{figure}

For the numerical simulation, we consider a honeycomb lattice with
length vectors $L_1\textbf{a}_1$ and $L_2\textbf{a}_2$ as shown in
the inset (a) of Figure~\ref{FigS:HoneycombFQHE}. Here
$\textbf{a}_1=(\sqrt{3},0)$ and
$\textbf{a}_2=(\frac{\sqrt{3}}{2},\frac{3}{2})$ are two primitive
vectors of the unit cell which includes two sites of the lattice.
The total number of sites is $N=2\times L_1\times L_2$, with
$L_1\times L_2$ unit cells. Note that the corresponding system width
$L_y=2L_2$, and system length $L_x=2L_1$. Unambiguously,
extrapolation from the data for $L_y\leq 20$ using
Eq.(\ref{Eq:EntropyDisk}) shows that we can get a nonzero constant
topological entanglement entropy $\gamma=0.349(5)$. This is equal to
$\ln(\sqrt{2})=0.347$ within the numerical error, showing that our
method can also be used to study chiral topological states as well.

\section{Transverse-field Ising model}
\begin{figure}
\centerline{\includegraphics[height=4.4in,width=3.4in] {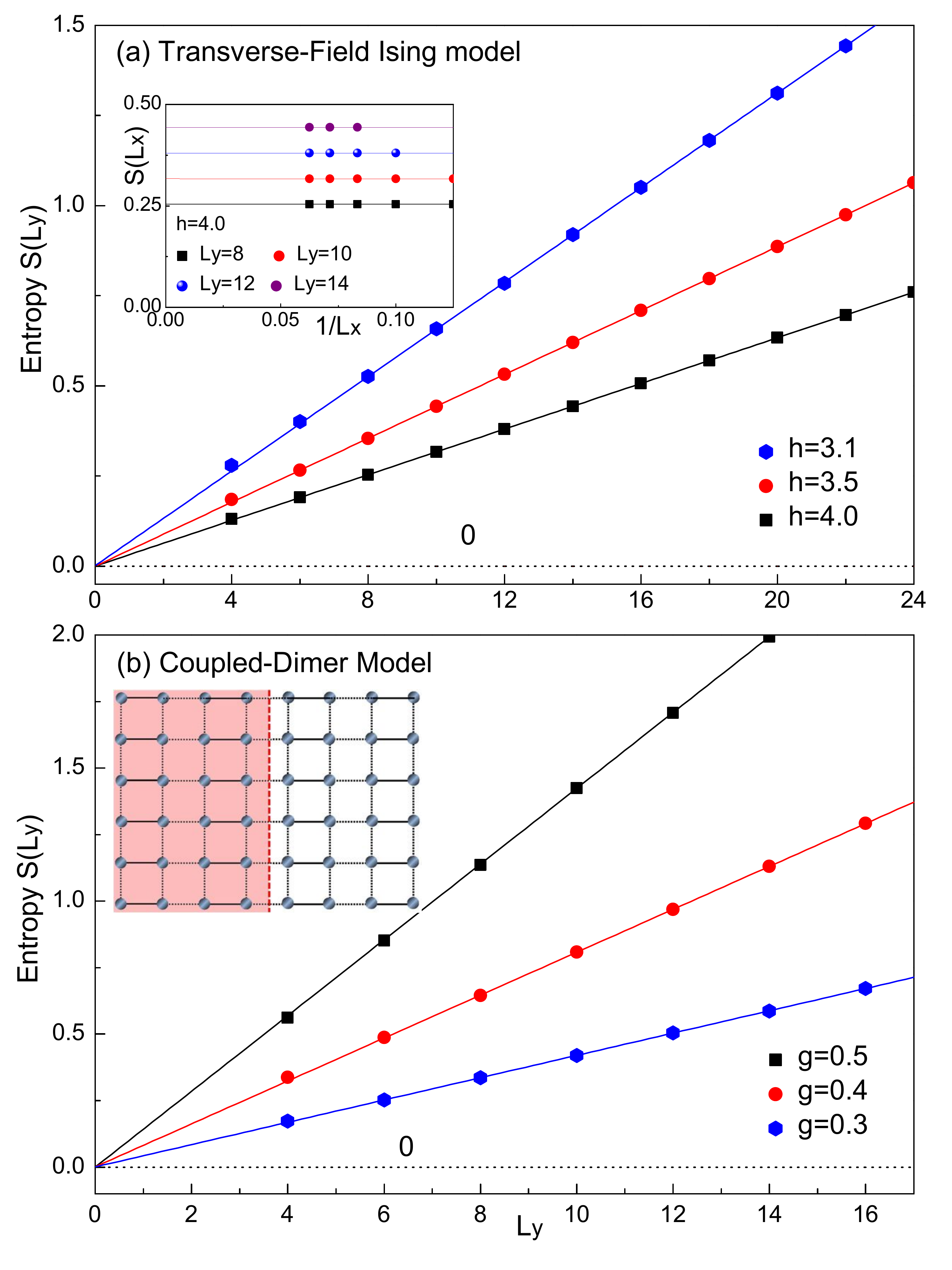}}%
\caption{The entanglement entropy of the transverse field quantum
Ising model and coupled spin-dimer model. (a) The entanglement
entropy $S(L_y)$ of the transverse field quantum Ising model in
Eq.(\ref{EqS:TransverseIsingModel}) with $L_y=4\sim 24$ at
$L_x=\infty$, in different magnetic field $h=3.1$, $3.5$ and $4.0$.
By fitting $S(L_y)=aL_y-\gamma$, we get $\gamma=0.0014(5)$,
$0.0004(4)$ and $0.0001(2)$, respectively. Inset: The entanglement
entropy $S(L_x)$ as a function of $L_x$ for different $L_y$ at
$h=4.0$. (b) The entanglement entropy $S(L_y)$ of the coupled
spin-dimer model in Eq.(\ref{EqS:CoupledDimerModel}) with $L_y=4\sim
16$ at $L_x=\infty$, at different coupling $g=0.5$, $0.4$ and $0.3$.
By fitting $S(L_y)=aL_y-\gamma$, we get $\gamma=0.006(6)$,
$0.002(1)$ and $0.0008(9)$, respectively. Inset: The coupled
spin-dimer model, with spin ($S=\frac{1}{2}$) on the sites, the $A$
links are shown as full lines, and the $B$ links as dashed lines.}
\label{FigS:IsingDimerModel}
\end{figure}

The models studied above have topologically ordered ground states,
for which our method indeed gives us non-zero topological
entanglement entropy with high accuracy. Now, we show that for a
topologically trivial phase, our method  unambiguously gives zero
topological entanglement entropy as well. To show this,
consider the well-known transverse field quantum Ising model%
\begin{eqnarray}
H=-\sum_{\langle ij\rangle}\sigma^z_i\sigma^z_j -
h\sum_i\sigma^x_i,\label{EqS:TransverseIsingModel}
\end{eqnarray}
where $\sigma^x_i$ and $\sigma^z_i$ are Pauli matrices on site $i$.
This model is known to have a topological trivial ground state in
all magnetic fields, and a second order phase transition at critical
field $h_c=3.044$\cite{Blote2002}.  As shown in
Figure~\ref{FigS:IsingDimerModel}(a), our method produces very
accurate results showing $\gamma=0$ even very close to the phase
transition point. For example, at $h=3.1$, we obtain
$\gamma=0.0014(5)$, which is zero within the numerical error,
despite the longish correlation length $\xi\sim 4$ for $<S^zS^z>$
and $\xi\sim 1$ for $<S^xS^x>$ in this case.

\section{Coupled spin-dimer model}
Another well-known model with a topologically trivial ground state is
the coupled spin-dimer model,
\begin{eqnarray}
H={\sum_{\langle ij\rangle\in
A}}\textbf{S}_i\cdot\textbf{S}_j+g{\sum_{\langle ij\rangle\in
B}}\textbf{S}_i\cdot\textbf{S}_j,\label{EqS:CoupledDimerModel}
\end{eqnarray}
where $\textbf{S}_i$ is the spin-$\frac{1}{2}$ operator on site $i$
on the square lattice shown in the inset of
Figure~\ref{FigS:IsingDimerModel}(b), with $A$ links forming
decoupled dimers while $B$ links couple the dimers. The ground state
of Eq.(\ref{EqS:CoupledDimerModel}) depends only on the
dimensionless coupling $g$. It is known that there is an gapped
dimerized phase for  $g<g_c=0.52$\cite{Gelfand1989,Sachdev2011}, at
which point a second order phase transition occurs. Unlike in the
transverse field quantum Ising model in
Eq.(\ref{EqS:TransverseIsingModel}), spin rotational symmetry is
preserved in this model, although the lattice translational symmetry
is explicitly broken. Unambiguously, as shown in
Figure~\ref{FigS:IsingDimerModel}(b), our method once more produces
very accurate results showing $\gamma=0$ even quite close to the
phase transition point. For example, $\gamma=0.006(6)$ at $g=0.50$,
which is zero within the numerical error.

\end{document}